\documentclass[aps,pra,twocolumn,showpacs,superscriptaddress]{revtex4-2}

\usepackage{amsmath, amssymb, bm}
\usepackage{graphicx}
\usepackage{physics}
\usepackage{hyperref}
\usepackage{tikz}
\usetikzlibrary{arrows.meta, decorations.pathmorphing, positioning}
\usepackage{float}

\begin{document}

\title{Persistence of Quantum Triality Relations in Open Qubit and Qutrit Systems}

\author{Pratidhwani Swain}\thanks{These authors contributed equally to this work.}
\affiliation{Department of Physics, Berhampur University, Berhampur, Odisha, India}
\email{ps.rs.phy@buodisha.edu.in}
\affiliation{Department of Physics, Nimapara Autonomous College, Bhubaneswar, Odisha, India} 

\author{Ramita Sarkar}\thanks{These authors contributed equally to this work.}
\email{ramitasarkar11@gmail.com}
\affiliation{Institute of Physics, Bhubaneswar, Odisha, India}
\affiliation{Department of Physical Sciences, Indian Institute of Science Education and Research (IISER) Kolkata, Mohanpur, India} 

\author{Sukanta K. Tripathy}
\affiliation{Department of Physics, Berhampur University, Berhampur, Odisha, India}
\email{skt.phy@buodisha.edu.in}

\author{Prasanta K. Panigrahi}
\affiliation{Department of Physical Sciences, Indian Institute of Science Education and Research (IISER) Kolkata, Mohanpur, India}
\affiliation{Center for Quantum Science and Technology, Siksha ’O’ Anusandhan University, Bhubaneswar, Odisha, India}
\email{pprasanta@iiserkol.ac.in}
\date{\today}

\begin{abstract}
We examine the complementarity among coherence (visibility), predictability, and entanglement for qubit and qutrit systems subjected to noisy quantum channels. Using the system--path entanglement framework, analytical expressions for all three quantities are derived for two- and three-slit interferometric setups. The study first establishes the validity of the triality relation in ideal conditions and then investigates its behavior under amplitude and phase damping. We find that amplitude damping redistributes coherence and population imbalance without violating complementarity, while phase damping reduces coherence but leaves predictability unchanged. These results demonstrate that the complementarity relation remains preserved even in open quantum systems, highlighting its robustness against decoherence and providing a unified analytical understanding of noisy quantum interferometry in low-dimensional systems.
\end{abstract}

\maketitle

Quantum entanglement is one of the cornerstone resources of quantum information theory \cite{dutta2019permutation, sarkar2021geometry,sarkar2021phase}. A composite system is said to be entangled when it cannot be represented as a direct product of its constituent subsystems. This inseparability is uniquely quantum in nature, with no counterpart in classical physics.
Surprisingly, however, correlations resembling entanglement can also emerge in structured optical fields \cite{azzini2020single, konrad2019quantum}. In such cases, the notion of classical entanglement or intra-system entanglement has been introduced to describe nonseparable couplings between different internal degrees of freedom of a single optical beam, for instance, its polarization and spatial modes.
Although rooted in classical physics, the mathematical formalism of classical entanglement \cite{paneru2020entanglement} mirrors that of its quantum counterpart. This striking similarity has opened the door to employing classical entanglement as a testbed for simulating quantum-inspired tasks, ranging from the implementation of quantum algorithms \cite{perez2018first}, quantum walks \cite{goyal2015implementation}, reconstruction of quantum channels \cite{ndagano2017characterizing}, quantum tomography \cite{xin2020quantum} and many more. \\
Wave-particle duality is the age-old interest in quantum mechanics \cite{bohr1928quantum}. The term quanton is defined for the quantum system which shows this dual nature. Among various approaches, two--path interference has served as a central testing ground for the study of this duality. Greenberger and Yasin introduced a simplified framework to address the issue of wave--particle duality. 
In their description, the notion of ``particle-like'' behavior is inferred only when the quanton has a higher probability of traveling through one path than the other. 
For such a case, they derived the following inequality~\cite{greenberger1988simultaneous}:
\begin{equation}
    \mathcal{P}^2 + \mathcal{V}^2 \leq 1,
    \label{inequality}
\end{equation}
where $\mathcal{P}$ represents the \textit{path predictability} (or simply, predictability), and 
$\mathcal{V}$ denotes the \textit{interference visibility}, which quantifies the wave-like nature of the quanton. 
In a typical two-path interference experiment, these two quantities capture the trade-off between the particle and wave characteristics. \\
Recent studies have unfolded the presence of different non-separable degrees of freedom in complementarity relations in optical systems \cite{qian2020turning, qian2016coherence}. The connection between two-particle visibility and concurrence has been drawn \cite{jakob2002quantitative}. Concurrence is one of the most important measures of entanglement, positive for entangled states and vanishing for separable states \cite{wootters1998entanglement, banerjee2020quantifying, bhaskara2017generalized}.  Complementarity
exists between the degree of polarization and  concurrence \cite{yugra2022constraints}. In recent studies, the triality relation has been derived including the polarization of the multi-slit interference experiment as follows \cite{qureshi2021predictability}:
\begin{equation}
    \mathcal{P}^2 + \mathcal{V}^2 +\mathcal{\varepsilon}^2= 1,
    \label{equality}
\end{equation}
Where, $\mathcal{P}$, $\mathcal{V}$ and $\mathcal{\varepsilon}$ are predictability, interference visibility(coherence) and entanglement. An exact triality relation has been established between which path,visibility and I concurrence following the conclusion reached by\cite{roy2024complementarity}.  \\
Qian and Eberly showed that classical non-deterministic fields  give a natural basis for entanglement and
Bell analyses in their work "entanglement is sometimes enough" \cite{qian2013entanglement}. In \cite{aiello2015quantum}, classical beam versions of well known quantum entangled states (GHZ, W, Noon) were obtained. \\
In this work, we investigate whether the triality relation among visibility, predictability, and entanglement remains valid under two common noise models---\textit{amplitude damping} and \textit{phase damping}---for both qubit and qutrit systems. 
In \textbf{section-\ref{sec:system_path}}, we describe the general framework of the \textit{system--path combination}, which enables the definition of entanglement even for a single quanton through its interaction with a path detector. 
\textbf{section-\ref{sec-3}} introduces the quantitative measures of \textit{coherence (visibility)}, \textit{predictability}, and \textit{entanglement}, which form the basis of the complementarity relation. 
\textbf{\ref{sec-4}} and \textbf{\ref{sec-5}} present detailed derivations of the triality condition for \textit{two-slit (qubit)} and \textit{three-slit (qutrit)} interferometric systems, respectively. 
In \textbf{\ref{sec-6}}, we analyze the influence of \textit{amplitude damping}, deriving analytical expressions for the modified visibility, predictability, and entanglement, and demonstrate that the triality relation persists. 
\textbf{\ref{7}} extends the study to \textit{phase damping}, where coherence decay alone is considered, again confirming the robustness of the complementarity relation. 
Finally, \textbf{\ref{8}} summarizes our findings and provides a concluding discussion on the persistence of complementarity under noisy quantum channels.

\section{System–Path Combination and Entanglement}
\label{sec:system_path}

In realistic interferometric scenarios, the quanton is not an isolated system but interacts with an ancillary degree of freedom, such as a path marker or detector. This composite description allows one to meaningfully discuss entanglement, even for single-particle systems. Following \cite{basso2020complete}, the joint system–detector state can be written as
\begin{equation}
	|\Psi\rangle = \sum_{i=1}^n c_i |i\rangle_{\mathrm{path}} \otimes |\phi_i\rangle_{\mathrm{det}},
\end{equation}
where $|i\rangle_{\mathrm{path}}$ denotes the path basis of the quanton and $|\phi_i\rangle_{\mathrm{det}}$ represents the correlated detector (or environment) states. The reduced density matrix of the quanton is obtained by tracing out the detector states:
\begin{equation}
	\rho = \mathrm{Tr}_{\mathrm{det}}(|\Psi\rangle\langle\Psi|).
\end{equation}
In this representation, the off-diagonal elements $\rho_{ij}$ encode the coherence between paths $i$ and $j$, while the diagonal elements $\rho_{ii}$ give the path populations. The entanglement measure $\varepsilon^2$ thus quantifies the correlations between the quanton and the path-detector system, consistent with the tripartite complementarity framework introduced in \cite{basso2020complete, qureshi2021predictability}.

\section{Coherence, Predictability, and Entanglement} \label{sec-3}
The measure of wave nature of light evidenced by visibility which is a picture of interference pattern observed. The more clarity of interference pattern leads more defined visibility or coherence. The mathematical framework of Coherence or visibility for a quanton passing through a multipath interferometer is given by \cite{roy2022coherence}:
\begin{equation}
	V^2 = \frac{n}{n-1} \sum_{i\neq j} |\rho_{ij}|^2
    \label{visibility}
\end{equation}
Similarly measurement of Predictability procures the particle nature of wave which in the context of multi-slit interferometer enables the amount of prediction of
path through which the photon is allowed to pass. Basically predictability overviews comprehensive idea of which path information and is formulated as \cite{roy2022coherence}
\begin{equation}
	P^2 = \sum_{i=1}^n \rho_{ii}^2 - \frac{1}{n-1} \sum_{i\neq j} \rho_{ii} \rho_{jj}
 \label{predictability}
\end{equation}
Finally, a normalized entanglement measure which emanates from the additive effect of both diagonal and off-diagonal elements of the density matrix as shown in \cite{roy2022coherence}
\begin{equation}
	\varepsilon^2 = \frac{n}{2(n-1)} \sum_{\mathrm{pairs}} E_{ij}^2,
\label{entanglement}
\end{equation}
where $E_{ij}$ denotes the concurrence between the quanton and the path detector. 

\section{Two-Slit Interferometer (Qubit System) and the Triality Relation}{\label{sec-4}}

For a \textbf{pure path state}, the quanton can be expressed as a coherent superposition of two paths,
\begin{equation}
|\psi\rangle = c_1 |\psi_1\rangle + c_2 |\psi_2\rangle.
\end{equation}
The corresponding density matrix is
\begin{equation}
\rho =
\begin{pmatrix}
|c_1|^2 & c_2 c_1^* \\
c_1 c_2^* & |c_2|^2
\end{pmatrix}.
\end{equation}
The visibility (or coherence) and predictability for this pure qubit state are obtained as
\begin{equation}
V^2 = 4|c_1|^2 |c_2|^2, \qquad P^2 = 1 - 4|c_1|^2 |c_2|^2,
\end{equation}
which clearly satisfy
\begin{equation}
V^2 + P^2 = 1.
\end{equation}
This equality demonstrates that the state is completely pure and that there is no entanglement with any detector degree of freedom. Hence, the path state can be written as a direct product with the detector, implying that the total system is \textit{separable} and non-entangled.

\vspace{1em}

When the quanton interacts with a path detector, the combined state is given by
\begin{equation}
|\Psi\rangle = c_1 |\psi_1\rangle |d_1\rangle + c_2 |\psi_2\rangle |d_2\rangle,
\end{equation}
where $|d_1\rangle$ and $|d_2\rangle$ represent the detector states correlated with each path. When the detector states are \emph{not identical}, i.e., $\langle d_1 | d_2 \rangle \neq 1$, the total state cannot be written as a simple product and becomes \textit{entangled}.

Tracing out the detector subsystem yields the reduced density matrix of the path degree of freedom,
\begin{equation}
\rho_r =
\begin{pmatrix}
|c_1|^2 & c_2 c_1^* \langle d_1 | d_2 \rangle \\
c_1 c_2^* \langle d_2 | d_1 \rangle & |c_2|^2
\end{pmatrix}.
\end{equation}
The off-diagonal elements are suppressed by the detector-state overlap:
\begin{equation}
\rho^{(r)}_{ij} = \rho_{ij} \langle d_i | d_j \rangle.
\end{equation}
The corresponding visibility and predictability are given by
\begin{equation}
V^2 = 4 |c_1|^2 |c_2|^2 |\langle d_1 | d_2 \rangle|^2, \qquad
P^2 = 1 - 4 |c_1|^2 |c_2|^2.
\end{equation}
The entanglement between the path and detector subsystems can then be expressed as
\begin{equation}
\varepsilon^2 = 4 |c_1|^2 |c_2|^2 \left( 1 - |\langle d_1 | d_2 \rangle|^2 \right).
\end{equation}

When the detector states are \textbf{parallel}, $|d_1\rangle = |d_2\rangle$, we have $|\langle d_1 | d_2 \rangle| = 1$, giving $\varepsilon^2 = 0$. In this limit, the combined system factorizes into a direct product of the path and detector parts, confirming that a pure path state corresponds to a separable (non-entangled) configuration.

If the detector states are not parallel, the reduced density matrix $\rho_r$ becomes a \textit{mixed state}, indicating the partial loss of coherence due to path–detector correlations.

\vspace{1em}

On the other hand, when the detector states are \textbf{orthogonal}, $\langle d_1 | d_2 \rangle = 0$, the system becomes \textit{maximally entangled}. In this case, the composite state
\begin{equation}
|\Psi\rangle = c_1 |\psi_1\rangle |d_1\rangle + c_2 |\psi_2\rangle |d_2\rangle
\end{equation}
reduces to the form of a \textbf{Bell-type entangled state}. For equal superposition amplitudes, $|c_1| = |c_2| = \frac{1}{\sqrt{2}}$, we obtain
\begin{equation}
|\Psi\rangle = \frac{1}{\sqrt{2}} \left( |\psi_1\rangle |d_1\rangle + |\psi_2\rangle |d_2\rangle \right),
\end{equation}
which represents a \textit{maximally entangled Bell state} between the path and detector subsystems. In this situation, the entanglement measure becomes $\varepsilon^2 = 1$, and both visibility and predictability vanish, $V^2 = P^2 = 0$.

\vspace{1em}

Therefore, the entanglement term $\varepsilon^2$ in the complementarity relation captures the degree of correlation between the path and detector states, extending the conventional wave–particle duality,
\begin{equation}
V^2 + P^2 = 1,
\end{equation}
to the complete \textbf{triality condition},
\begin{equation}
V^2 + P^2 + \varepsilon^2 = 1.
\end{equation}
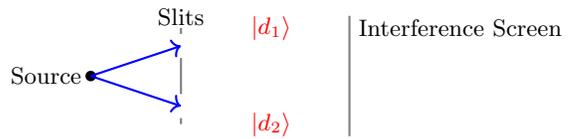
\begin{figure}[ht]
\centering
\begin{tikzpicture}[scale=0.8, thick] 
  \filldraw[black] (-3,0) circle (0.07);
  \node[left] at (-3,0) {Source};

  \draw[gray, thick] (-1.5,0.8) -- (-1.5,-0.8);
  \draw[white, line width=4pt] (-1.5,0.3) -- (-1.5,0.7);
  \draw[white, line width=4pt] (-1.5,-0.7) -- (-1.5,-0.3);
  \node[above] at (-1.5,0.7) {Slits};

  \draw[->,blue,thick] (-3,0) -- (-1.5,0.5);
  \draw[->,blue,thick] (-3,0) -- (-1.5,-0.5);

  \draw[gray, thick] (1.3,1) -- (1.3,-1);
  \node[right] at (1.3,0.8) {Interference Screen};

  \node[red] at (0,0.8) {$|d_1\rangle$};
  \node[red] at (0,-0.8) {$|d_2\rangle$};

\end{tikzpicture}
\caption{Two-slit interferometer with path detector states $|d_1\rangle$ and $|d_2\rangle$.}
\label{fig:twoslit_pra}
\end{figure}

\subsection*{Alternative Pauli–Matrix Representation}

For completeness, the qubit density operator can also be expressed in the Pauli matrix representation as
\begin{equation}
\rho = \rho_0 I + \boldsymbol{\rho} \cdot \boldsymbol{\sigma}
= 
\begin{pmatrix}
\rho_0 + \rho_3 & \rho_1 - i\rho_2 \\
\rho_1 + i\rho_2 & \rho_0 - \rho_3
\end{pmatrix},
\end{equation}
where $\boldsymbol{\sigma} = (\sigma_x, \sigma_y, \sigma_z)$ are the Pauli matrices and $\rho_0 = \tfrac{1}{2}$ ensures $\mathrm{Tr}(\rho)=1$.
In this form, the off-diagonal elements encode the coherence terms $\rho_{12}$ and $\rho_{21}$, while the population imbalance appears through $\rho_3$.

Using this representation, the visibility and predictability can be rewritten as
\begin{equation}
V^2 = 4|\rho_{12}|^2 = 4(\rho_1^2 + \rho_2^2), \qquad
P^2 = (\rho_{11} - \rho_{22})^2 = 4\rho_3^2.
\end{equation}
The entanglement measure, obtained from the reduced density matrix after tracing out the detector degrees of freedom, becomes
\begin{equation}
\varepsilon^2 = 4\rho_0^2 - V^2 - P^2.
\end{equation}
Since $\rho_0 = \tfrac{1}{2}$ for a normalized two-level system, we obtain
\begin{equation}
V^2 + P^2 + \varepsilon^2 = 1,
\end{equation}
which is identical to the triality condition derived previously. 

This alternative formulation demonstrates that the same complementarity relation holds naturally in the Bloch (Pauli) representation of the qubit state, confirming the consistency of the approach across both path–detector and algebraic frameworks.

\section{Three-Slit (Qutrit) Interferometer and Extension of the Triality Relation}\label{sec-5}

Having established the triality relation for the two-path (qubit) interferometer and verified its consistency in both the path–detector and Pauli-matrix representations, we now extend the analysis to a three-path (qutrit) system. The inclusion of a third path enriches the interference structure and provides a natural framework for exploring complementarity in higher-dimensional Hilbert spaces, where coherence, predictability, and entanglement coexist in more intricate forms.

For a pure qutrit state, the quanton can be expressed as
\begin{equation}
|\psi\rangle = c_1|\psi_1\rangle + c_2|\psi_2\rangle + c_3|\psi_3\rangle,
\end{equation}
with the corresponding density matrix
\begin{equation}
\rho =
\begin{pmatrix}
|c_1|^2 & c_1c_2^* & c_1c_3^*\\
c_2c_1^* & |c_2|^2 & c_2c_3^*\\
c_3c_1^* & c_3c_2^* & |c_3|^2
\end{pmatrix}.
\end{equation}

From this, the visibility and predictability for the pure qutrit state are obtained as
\begin{align}
V^2 &= 3(|c_1|^2|c_2|^2 + |c_2|^2|c_3|^2 + |c_3|^2|c_1|^2),\\
P^2 &= 1 - 3(|c_1|^2|c_2|^2 + |c_2|^2|c_3|^2 + |c_3|^2|c_1|^2),
\end{align}
which satisfy the duality relation $V^2 + P^2 = 1$ in the absence of entanglement.

When a path–detector correlation is introduced, the combined state becomes
\begin{equation}
|\Psi\rangle = c_1|\psi_1\rangle|d_1\rangle + c_2|\psi_2\rangle|d_2\rangle + c_3|\psi_3\rangle|d_3\rangle,
\end{equation}

\begin{figure}[h]
\centering
\begin{tikzpicture}[scale=1.1, thick, >=Stealth]

\filldraw[black] (-3,0) circle (0.07);
\node[left] at (-3,0) {Source};

\draw[gray, thick] (-1.5,1.3) -- (-1.5,-1.3);
\foreach \y in {0.9,0.0,-0.9}
  \draw[white, line width=4pt] (-1.5,\y-0.15) -- (-1.5,\y+0.15);
\node[above] at (-1.5,1.3) {Three Slits};

\foreach \y in {0.9,0.0,-0.9}
  \draw[->,blue,thick] (-3,0) -- (-1.5,\y);

\node[red] at (-0.3,0.9) {$|d_1\rangle$};
\node[red] at (-0.3,0.0) {$|d_2\rangle$};
\node[red] at (-0.3,-0.9) {$|d_3\rangle$};

\draw[gray, thick] (2,1.3) -- (2,-1.3);
\node[right] at (2,1.1) {Screen};

\end{tikzpicture}
\caption{Three-slit (qutrit) interferometer showing path--detector correlations $|d_1\rangle$, $|d_2\rangle$, and $|d_3\rangle$.}
\end{figure}
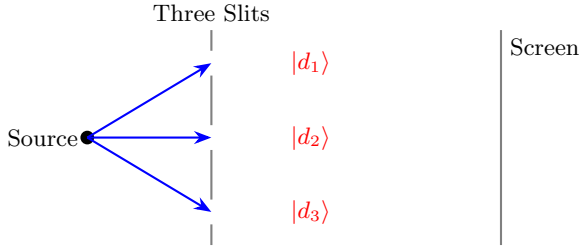

where $|d_i\rangle$ represent the detector states associated with each path. Tracing out the detector degrees of freedom yields the reduced density matrix
\begin{equation}
\rho_r =
\begin{pmatrix}
|c_1|^2 & c_1c_2^*\langle d_1|d_2\rangle & c_1c_3^*\langle d_1|d_3\rangle\\
c_2c_1^*\langle d_2|d_1\rangle & |c_2|^2 & c_2c_3^*\langle d_2|d_3\rangle\\
c_3c_1^*\langle d_3|d_1\rangle & c_3c_2^*\langle d_3|d_2\rangle & |c_3|^2
\end{pmatrix}.
\end{equation}

The corresponding measures of visibility, predictability, and entanglement become
\begin{align}
V^2 &= 3\Big( 
       |c_1|^2 |c_2|^2 |\langle d_1 | d_2 \rangle|^2 
       + |c_2|^2 |c_3|^2 |\langle d_2 | d_3 \rangle|^2 \nonumber\\
&\qquad + |c_3|^2 |c_1|^2 |\langle d_3 | d_1 \rangle|^2 
       \Big), \\
P^2 &= 1 - 3 \Big( |c_1|^2 |c_2|^2 
       + |c_2|^2 |c_3|^2 
       + |c_3|^2 |c_1|^2 \Big), \\
\varepsilon^2 &= 3 |c_1|^2 |c_2|^2 \Big(1 - |\langle d_1 | d_2 \rangle|^2 \Big) \nonumber\\
&\quad + 3 |c_2|^2 |c_3|^2 \Big(1 - |\langle d_2 | d_3 \rangle|^2 \Big) \nonumber\\
&\quad + 3 |c_3|^2 |c_1|^2 \Big(1 - |\langle d_3 | d_1 \rangle|^2 \Big).
\end{align}

These expressions together satisfy the generalized triality condition
\begin{equation}
V^2 + P^2 + \varepsilon^2 = 1,
\end{equation}
confirming that the complementarity relation extends naturally to a three-path interferometer.

To connect with the algebraic (Bloch-like) representation, the density operator for the qutrit can be expanded in terms of the Gell–Mann matrices $\lambda_i$ as
\begin{equation}
\rho = \frac{I_3}{3} + \frac{1}{\sqrt{3}}\sum_{i=1}^{8} S_i\lambda_i,
\end{equation}
leading to
\begin{align}
V^2 &= S_1^2 + S_2^2 + S_4^2 + S_5^2 + S_6^2 + S_7^2,\\
P^2 &= S_3^2 + S_8^2.
\end{align}
Consequently, the entanglement can be written as
\begin{equation}
\varepsilon^2 = 1 - (S_1^2 + S_2^2 + S_3^2 + S_4^2 + S_5^2 + S_6^2 + S_7^2 + S_8^2).
\end{equation}

Substituting the expressions for $V^2$ and $P^2$ immediately yields
\begin{equation*}
V^2 + P^2 + \varepsilon^2 = 1.
\end{equation*}

Thus, the three-slit (qutrit) interferometer satisfies the same fundamental complementarity condition as the qubit case. The introduction of the third path enhances the richness of the interference and entanglement structure while preserving the unity of the triality relation, confirming its universality across different Hilbert-space dimensions.

\section{Complementarity under Amplitude Damping Channels}\label{sec-6}

\subsection{Qubit system under amplitude damping}
The amplitude damping channel for a qubit is represented by the Kraus operators
\begin{equation}
	E_0 =
	\begin{pmatrix}
		1 & 0 \\
		0 & \sqrt{1-\gamma_A}
	\end{pmatrix}, \quad
	E_1 =
	\begin{pmatrix}
		0 & \sqrt{\gamma_A} \\
		0 & 0
	\end{pmatrix},
\end{equation}
where $0 \le \gamma_A \le 1$ is the damping parameter. The evolved density matrix is
\begin{equation}
	\rho' = \sum_{k=0}^1 E_k \rho E_k^\dagger =
	\begin{pmatrix}
		\rho_{11} + \gamma_A \rho_{22} & \sqrt{1-\gamma_A}\rho_{12} \\
		\sqrt{1-\gamma_A}\rho_{21} & (1-\gamma_A)\rho_{22}
	\end{pmatrix}.
\end{equation}
The new predictability becomes
\begin{equation}
	P'^2 = (\rho'_{11})^2 + (\rho'_{22})^2 - 2\rho'_{11}\rho'_{22} = \big(\rho_{11} - \rho_{22} + 2\gamma_A \rho_{22}\big)^2,
\end{equation}
and the visibility reduces to
\begin{equation}
	V'^2 = 4|\rho'_{12}|^2 = (1-\gamma_A) V^2.
\end{equation}
entanglement in this case
\begin{align}
{\epsilon'}^{2}=E_{12}'^{\,2} 
&= 4(1-\gamma_A)\big(\rho_{11}\rho_{22}-|\rho_{12}|^2\big) \notag \\[6pt]
&\quad + 4\gamma_A(1-\gamma_A)\rho_{22}^2 \notag \\[6pt]
&= (1-\gamma_A)\,E_{12}^{\,2} 
   + 4\gamma_A(1-\gamma_A)\rho_{22}^2.
\end{align}

\begin{equation}
 P'^{2}+V'^{2}+{\epsilon'}^{2}
= (\rho_{11}+\rho_{22})^{2} 
=1   
\end{equation}
We found that triality relation holds for the amplitude damping case.
\subsection{Qutrit system under amplitude damping}
For the qutrit case, we consider a cascade decay $|2\rangle \to |1\rangle \to |0\rangle$ with rates $\gamma_2$ and $\gamma_1$. The Kraus operators are
\begin{equation}
\begin{aligned}
E_0 &= |0\rangle\langle0| + \sqrt{1-\gamma_1}\,|1\rangle\langle1|
      + \sqrt{1-\gamma_2}\,|2\rangle\langle2|,\\
E_1 &= \sqrt{\gamma_1}\,|0\rangle\langle1|,\\
E_2 &= \sqrt{\gamma_2}\,|1\rangle\langle2|.
\end{aligned}
\label{eq:kraus}
\end{equation}

The evolved density matrix is $\rho' = \sum_{k=0}^2 E_k \rho E_k^\dagger$ with updated populations
\begin{align}
	\rho'_{00} &= \rho_{00} + \gamma_1 \rho_{11} + \gamma_1 \gamma_2 \rho_{22}, \\
	\rho'_{11} &= (1-\gamma_1)\rho_{11} + \gamma_2(1-\gamma_1)\rho_{22}, \\
	\rho'_{22} &= (1-\gamma_2)\rho_{22}.
\end{align}
Off-diagonal elements are damped as
\begin{equation}
	\rho'_{ij} = \sqrt{(1-\gamma_i)(1-\gamma_j)}\rho_{ij} \quad (i\neq j).
\end{equation}

\begin{equation}
\begin{aligned}
s'_{1,2} &= \sqrt{1-\gamma_1}\; s_{1,2}, \\
s'_{4,5} &= \sqrt{1-\gamma_2}\; s_{4,5} \\
s'_{6,7} &= \sqrt{(1-\gamma_1)(1-\gamma_2)}\; s_{6,7}\\
s'_8 &= (1-\gamma_2)\; s_8 + \frac{\gamma_2}{2}, \\
s'_3 &= (1-\gamma_1)\; s_3 + \frac{\sqrt{3}}{3}\,\gamma_1(1-\gamma_2)\; s_8 \; 
\end{aligned}
\label{eq:transforms}
\end{equation}

Visibility
\[
V'^2 = (s_1')^2 + (s_2')^2 + (s_4')^2 + (s_5')^2 + (s_6')^2 + (s_7')^2
\]
Predictability
\[
P'^2 = (s_3')^2 + (s_8')^2
\]

 Entanglement
\begin{equation}
\begin{aligned}
\epsilon'^2 = 1 - \Big[ &(s_1')^2 + (s_2')^2 + (s_3')^2 + (s_4')^2  \\
                 &+ (s_5')^2 + (s_6')^2 + (s_7')^2 + (s_8')^2 \Big].
\end{aligned}
\label{eq:Eprime2}
\end{equation}

We also found that 
\[
V'^2 + P'^2 + \epsilon'^2 = 1
\]

\subsection{Persistence of Triality}
Notably, the triality relation $V'^2 + P'^2 + \varepsilon'^2 = 1$
remains valid under amplitude damping. The channel redistributes the contributions of coherence, predictability, and entanglement without breaking their sum rule.
\section{Complementarity under Phase Damping (Dephasing) Channels}\label{7}
\subsection{Qubit system under phase damping (dephasing)}

The phase damping channel for a qubit is represented by the Kraus operators
\begin{equation}
E_0 =
\begin{pmatrix}
1 & 0 \\
0 & \sqrt{1-\gamma_p}
\end{pmatrix}, \qquad
E_1 =
\begin{pmatrix}
0 & 0 \\
0 & \sqrt{\gamma_p}
\end{pmatrix},
\label{eq:pd_kraus}
\end{equation}
where $0 \leq \gamma_p \leq 1$ is the dephasing parameter. The evolved density matrix is
\begin{equation}
\rho' = \sum_{k=0}^1 E_k \rho E_k^\dagger =
\begin{pmatrix}
\rho_{11} & \sqrt{1-\gamma_p}\,\rho_{12} \\
\sqrt{1-\gamma_p}\,\rho_{21} & \rho_{22}
\end{pmatrix}.
\label{eq:pd_rho}
\end{equation}

The predictability is unaffected by phase damping,
\begin{equation}
P'^2 = (\rho_{11} - \rho_{22})^2 = P^2,
\label{eq:pd_predictability}
\end{equation}
while the visibility reduces according to
\begin{equation}
V'^2 = 4|\rho'_{12}|^2 = (1-\gamma_p)V^2.
\label{eq:pd_visibility}
\end{equation}

\begin{equation}
 {\epsilon'}^{2}=E_{12}'^{\,2} =
E_{12}^{\,2} 
   + 4\gamma_p|\rho_{12}|^2.   
\end{equation}

Here also we see, 
\begin{equation}
 P'^{2}+V'^{2}+{\epsilon'}^{2}
=1   
\end{equation}
Thus, phase damping diminishes only the coherence, leaving the population imbalance unchanged, and redistributes the balance between visibility and entanglement without violating the complementarity relation.

\subsection{Qutrit system under phase damping (dephasing)}

The phase damping (or pure dephasing) channel for a three-level system 
is represented by the following set of Kraus operators~\cite{nielsen2010quantum,siudzinska2017generalized,lidar2014review}:
\begin{equation}
\begin{aligned}
E_0 &= 
\sqrt{\frac{1 + 2\sqrt{1 - \gamma_p}}{3}}
\begin{pmatrix}
1 & 0 & 0 \\[2pt]
0 & 1 & 0 \\[2pt]
0 & 0 & 1
\end{pmatrix}, \\[8pt]
E_1 &= 
\sqrt{\frac{1 - \sqrt{1 - \gamma_p}}{3}}
\begin{pmatrix}
1 & 0 & 0 \\[2pt]
0 & \omega & 0 \\[2pt]
0 & 0 & \omega^2
\end{pmatrix}, \\[8pt]
E_2 &= 
\sqrt{\frac{1 - \sqrt{1 - \gamma_p}}{3}}
\begin{pmatrix}
1 & 0 & 0 \\[2pt]
0 & \omega^2 & 0 \\[2pt]
0 & 0 & \omega
\end{pmatrix}.
\end{aligned}
\label{eq:kraus_qutrit}
\end{equation}

where $\omega = e^{2\pi i/3}$ and $0 \leq \gamma_p \leq 1$ denotes the dephasing parameter.
These operators satisfy the trace-preserving completeness relation
\begin{equation}
E_0^\dagger E_0 + E_1^\dagger E_1 + E_2^\dagger E_2 = I_3,
\end{equation}
ensuring a completely positive, trace-preserving map.

Under this channel, the diagonal elements of the density matrix remain 
unchanged, while the off-diagonal elements are attenuated:
\begin{equation}
\rho'_{ii} = \rho_{ii}, \qquad 
\rho'_{ij} = \sqrt{1 - \gamma_p}\,\rho_{ij} \ \ (i \neq j).
\end{equation}
In the Gell--Mann representation, this leads to the transformation
\begin{equation}
s'_{1,2,4,5,6,7} = \sqrt{1 - \gamma_p}\, s_{1,2,4,5,6,7}, 
\quad s'_3 = s_3, \quad s'_8 = s_8.
\end{equation}

Consequently, the modified quantities become
\begin{equation}
V'^2 = (1 - \gamma_p)V^2, \quad P'^2 = P^2, \quad 
\varepsilon'^2 = 1 - P^2 - (1 - \gamma_p)V^2.
\end{equation}
Therefore, the complementarity (triality) relation remains valid under 
qutrit phase damping:
\begin{equation*}
V'^2 + P'^2 + \varepsilon'^2 = 1.
\end{equation*}

\section{Conclusions}\label{8}
In this work, we have carried out a detailed analytical investigation of the complementarity relation among visibility, predictability, and entanglement for two- and three-path interferometric systems. Using the system--path framework, we derived closed-form expressions for these quantities in both qubit and qutrit settings and confirmed that the triality condition $V^{2} + P^{2} + \varepsilon^{2} = 1$ 
holds exactly for pure, isolated systems. We further examined how this relation behaves under two widely studied noisy quantum 
channels: amplitude damping and phase damping. For both qubit and qutrit systems, our results show that although decoherence redistributes coherence, population 
imbalance, and system--environment correlations in distinct ways, the triality relation remains strictly preserved. Amplitude damping affects both coherence and 
populations, while phase damping suppresses only the off-diagonal elements; yet in both cases the sum of visibility, predictability, and entanglement continues to satisfy the complementarity equality.

These findings demonstrate that the triality relation is structurally robust, not only in closed interferometric systems but also under realistic open-system dynamics. The unified treatment in two- and three-dimensional Hilbert spaces provides a consistent framework for understanding how coherence and entanglement co-evolve under noise. The analysis presented here may support future investigations of complementarity in higher-dimensional systems, non-Markovian evolution, and experimentally relevant photonic, atomic, or classical–quantum analog platforms.

\bibliographystyle{apsrev4-2}
\bibliography{library}
\appendix
\section{Illustrative Example: Evolution of $V'^2$, $P'^2$, and $\varepsilon'^2$ under amplitude damping}

To visualize the effect of amplitude damping, we consider simple maximally coherent initial states for qubit and qutrit systems and track the evolution of coherence, predictability, and entanglement.

\subsection{Qubit Example}
We take the initial qubit state as the pure equal superposition
\begin{equation}
\rho = 
\begin{pmatrix}
\tfrac{1}{2} & \tfrac{1}{2} \\
\tfrac{1}{2} & \tfrac{1}{2}
\end{pmatrix},
\label{eq:A1}
\end{equation}
for which the initial values are $V^2 = 1$, $P^2 = 0$, and $\varepsilon^2 = 0$.
After amplitude damping, the evolved state becomes
\begin{equation}
\rho' =
\begin{pmatrix}
\tfrac{1}{2} + \tfrac{\gamma}{2} & \tfrac{1}{2}\sqrt{1-\gamma} \\
\tfrac{1}{2}\sqrt{1-\gamma} & \tfrac{1}{2}(1-\gamma)
\end{pmatrix}.
\end{equation}
Now we obtain
\begin{equation}
\begin{aligned}
V'^2 &= 1 - \gamma,\\
P'^2 &= \gamma^2,\\
\varepsilon'^2 &= 1 - V'^2 - P'^2 = \gamma - \gamma^2.
\end{aligned}
\end{equation}

\subsection{Qutrit Example}
For the qutrit, we take the maximally coherent state
\begin{equation}
\rho = \tfrac{1}{3}
\begin{pmatrix}
1 & 1 & 1 \\
1 & 1 & 1 \\
1 & 1 & 1
\end{pmatrix},
\label{eq:qt}
\end{equation}
which yields $V^2 = 1$, $P^2 = 0$, and $\varepsilon^2 = 0$. 
Under cascade amplitude damping with equal decay rates $\gamma_1 = \gamma_2 = \gamma$, the diagonal elements evolve as
\begin{equation}
\begin{aligned}
\rho'_{00} &= \tfrac{1}{3}(1 + \gamma + \gamma^2),\\
\rho'_{11} &= \tfrac{1}{3}(1 - \gamma^2),\\
\rho'_{22} &= \tfrac{1}{3}(1 - \gamma),
\end{aligned}
\end{equation}
while all off-diagonal coherences are multiplied by $(1-\gamma)$. Thus
\begin{equation}
V'^2 = (1-\gamma)^2.
\end{equation}
The updated predictability, using Eq.~(2), simplifies to
\begin{equation}
P'^2 = \tfrac{1}{2}\big[(\rho'_{00}-\rho'_{11})^2 
+ (\rho'_{11}-\rho'_{22})^2 
+ (\rho'_{22}-\rho'_{00})^2\big],
\end{equation}
and entanglement again follows from the triality condition,
\begin{equation}
\varepsilon'^2 = 1 - V'^2 - P'^2.
\end{equation}


\section{ Illustrative Examples — Evolution of \(V'^{2}\), \(P'^{2}\), and \(\varepsilon'^{2}\) under Phase Damping}

We now analyze the effect of phase damping (pure dephasing) on the same qubit and qutrit systems. The relevant Kraus operators and transformations correspond to Eqs.~(52)–(62) in the main text.

\subsection*{B.1 Qubit Example}

We use phase-damping channel for qubit to the maximally coherent state (\ref{eq:A1}), the evolved density matrix becomes
\begin{equation}
\rho' =
\begin{pmatrix}
\frac{1}{2} & \tfrac{1}{2}\sqrt{1-\gamma_{p}}\\[4pt]
\tfrac{1}{2}\sqrt{1-\gamma_{p}} & \frac{1}{2}
\end{pmatrix}.
\label{eq:B2}
\end{equation}

The off-diagonal terms decay by \(\sqrt{1-\gamma_{p}}\), while the populations remain constant. Thus,
\begin{equation}
V'^{2}=1-\gamma_{p}, \qquad
P'^{2}=0, \qquad
\varepsilon'^{2}=\gamma_{p},
\label{eq:B3}
\end{equation}
satisfying
\begin{equation}
V'^{2}+P'^{2}+\varepsilon'^{2}=1.
\label{eq:B4}
\end{equation}

\subsection*{B.2 Qutrit Example}

For the three-level system, applying the dephasing channel to the maximally coherent qutrit state \ref{eq:qt} yields
\begin{equation}
\rho' = \frac{1}{3}
\begin{pmatrix}
1 & \sqrt{1-\gamma_{p}} & \sqrt{1-\gamma_{p}}\\[4pt]
\sqrt{1-\gamma_{p}} & 1 & \sqrt{1-\gamma_{p}}\\[4pt]
\sqrt{1-\gamma_{p}} & \sqrt{1-\gamma_{p}} & 1
\end{pmatrix}.
\label{eq:B6}
\end{equation}

Since only the coherences are attenuated,
\begin{equation}
V'^{2}=1-\gamma_{p}, \qquad
P'^{2}=0, \qquad
\varepsilon'^{2}=\gamma_{p},
\label{eq:B7}
\end{equation}
so that, $V'^{2}+P'^{2}+\varepsilon'^{2}=1$ holds.

\begin{figure}[h]
\centering
\includegraphics[width=1\linewidth]{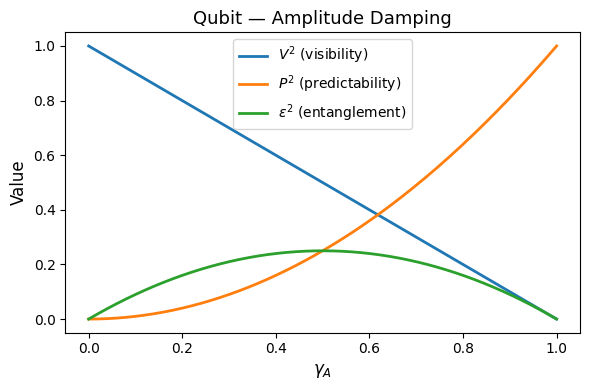}
\includegraphics[width=1\linewidth]{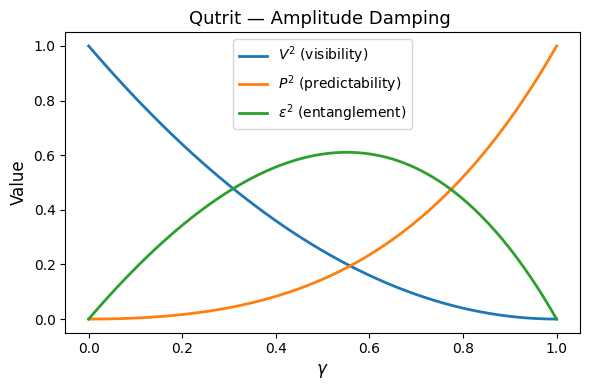}
\caption{(a) Qubit and (b) qutrit visibility, predictability, and entanglement under amplitude damping.}
\label{fig:combined}
\end{figure}

\begin{figure}[h]
\centering
\includegraphics[width=1\linewidth]{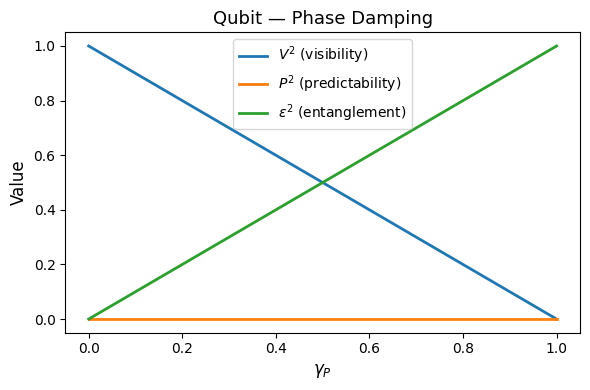}
\includegraphics[width=1\linewidth]{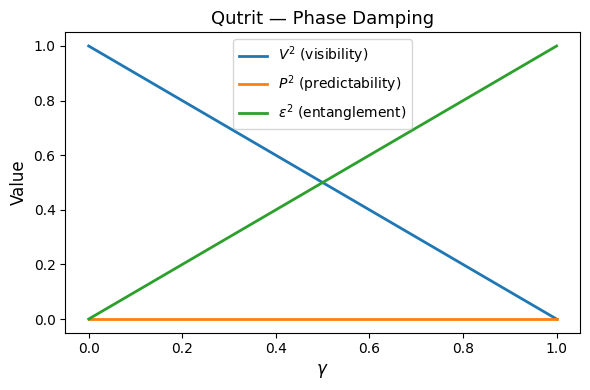}
\caption{(a) Qubit and (b) qutrit visibility, predictability, and entanglement under phase damping.}
\label{fig:combined2}
\end{figure}

\end{document}